\documentclass[a4paper,twoside]{article}
\usepackage{times}
\usepackage{xspace}
\usepackage{subfig}
\usepackage{epsfig}
\usepackage{caption}
\usepackage[table]{xcolor}
\usepackage{bbm}
\usepackage{array,longtable}
\usepackage{balance}
\usepackage{graphicx}
\usepackage{SCITEPRESS} 
\usepackage{crossreference}
\usepackage{tikz}
\usepackage{booktabs}
\usepackage{eurosym}
\usepackage{array}
\usepackage{adjustbox}
\usepackage{lipsum}
\usepackage{multirow}
\usepackage{hyphenat}

\usepackage{listings}

\usepackage{footnote}
\makesavenoteenv{tabular}
\makesavenoteenv{table}
\usepackage{color}
\usepackage{hhline}
\usepackage{booktabs}
\usepackage{multirow}
\usepackage{SCITEPRESS}     % Please add other packages that you may need BEFORE the SCITEPRESS.sty package.

\usepackage{hyperref}
\hypersetup{colorlinks = true, citecolor= red}
\newcommand{\Chi}[2]{%
  \csname CJK*\endcsname{UTF8}{zhsong}%
    \CJKchar{#1}{#2}%
  \csname endCJK*\endcsname
}

\usepackage{etoolbox}

\makeatletter
\preto\lstlisting{\def\@captype{table}}
\makeatother
\usepackage{pifont}

\renewcommand{\footnotesize}{\fontsize{9}{10}\selectfont}

\usepackage{flushend}

\usepackage{amssymb}% http://ctan.org/pkg/amssymb
\usepackage{pifont}% http://ctan.org/pkg/pifont

\usepackage{etoolbox}
\makeatletter
\preto\lstlisting{\def\@captype{table}}
\makeatother

\renewcommand{\footnotesize}{\fontsize{9}{10}\selectfont}

\usepackage{todonotes, xpatch,hyperref}

\let\oldsec=\section
\renewcommand*{\section}{\secdef{\Sec}{\SecS}}
\newcommand\SecS[1]{\oldsec*{#1}}%
\newcommand\Sec[2][]{\oldsec[\texorpdfstring{#1}{#1}]{#2}}%

\makeatletter
\xapptocmd{\Sec}{\addtocontents{tdo}{\protect\todoline{\thesection}{#1}{}}}{}{}
\newcommand{\todoline}[1]{\@ifnextchar\Endoftdo{}{\@todoline{#1}}}
\newcommand{\@todoline}[3]{%
	\@ifnextchar\todoline
	{}
	{\contentsline{section}{\numberline{#1}#2}{#3}{}{}}%
}
\let\l@todo\l@subsection
\newcommand{\Endoftdo}{}

\AtEndDocument{\addtocontents{tdo}{\string\Endoftdo}}
\usepackage{flushend}

\graphicspath{ {images/paperImages/} }

\usepackage{array,longtable}
\usepackage{etoolbox}

\usepackage{flushend}

\usepackage{subfiles}

\newcolumntype{L}[1]{>{\raggedright\let\newline\\\arraybackslash\hspace{0pt}}m{#1}}
\newcolumntype{C}[1]{>{\centering\let\newline\\\arraybackslash\hspace{0pt}}m{#1}}
\newcolumntype{R}[1]{>{\raggedleft\let\newline\\\arraybackslash\hspace{0pt}}m{#1}}

\begin{document}

\title{DaDiDroid: An Obfuscation Resilient Tool for Detecting Android Malware via Weighted Directed Call Graph Modelling}
\author{\authorname{Muhammad Ikram\sup{1, 3}, Pierrick Beaume\sup{2}, and Mohamed Ali Kaafar\sup{1}\sup{2}}
\affiliation{\sup{1}Macquarie University, \sup{2}Data61 CSIRO, \sup{3}University of Michigan}
\email{\{Muhammad.Ikram, Dali.Kaafar\}@mq.edu.au, Pierrick.Beaume@data61.csiro.au}
}
\abstract{
With the number of new mobile malware instances increasing by over 50\% annually since 2012 \cite{mcafeeReport}, malware embedding in mobile apps is arguably one of the most serious security issues mobile platforms are exposed to. While obfuscation techniques are successfully used to protect the intellectual property of apps' developers, they are unfortunately also often used by cybercriminals to hide malicious content inside mobile apps and to deceive malware detection tools. As a consequence, most of mobile malware detection approaches fail in differentiating between benign and obfuscated malicious apps.      
We examine the graph features of mobile apps code by building weighted directed graphs of the API calls, and verify that malicious apps often share structural similarities that can be used to differentiate them from benign apps, even under a heavily ``polluted'' training set where a large majority of the apps are obfuscated.    
We present DaDiDroid  an Android malware app detection tool that leverages features of the weighted directed graphs of API calls to detect the presence of malware code in (obfuscated) Android apps. We show that DaDiDroid significantly outperforms MaMaDroid~\cite{MaMaDroid}, a recently proposed malware detection tool that has been proven very efficient in detecting malware in a clean non-obfuscated environment.  
We evaluate DaDiDroid's accuracy and robustness against several evasion techniques using various datasets for a total of 43,262 benign and 20,431 malware apps. We show that DaDiDroid correctly labels up to 96\% of Android malware samples, while achieving an 91\% accuracy with an exclusive use of a training set of obfuscated apps. 
\vspace{-0.75cm}
}

\maketitle %\normalsize \setcounter{footnote}{0} %\vfill
\vspace{-2.75cm}
\section{\uppercase{Introduction}}
\label{sec:intro}
\vspace{-0.25cm}

In recent years, Android OS and mobile applications (apps in short) have experienced an exponential growth with over 3 Million apps on Google Play~\cite{gPlay} in 2017. This popularity naturally attracted malware developers leading to a continuous 50\% yearly increase of the number of malware apps for over five years~\cite{mcafeeReport}.   
Several research studies (e.g.,~\cite{ikram2016analysis, AppFence, taintDroid})  
focused on dynamic runtime-network analysis to detect potential malicious behaviour of Android apps. Another line of research (e.g., \cite{StaticPrivacyLeaks, arp2014drebin, chen2016stormdroid, aafer2013droidapiminer}) aimed to reduce the resources required to perform dynamic analysis by adopting static code analysis approaches~\cite{ikram2016analysis, ikramfirst}. DroidAPIMiner~\cite{aafer2013droidapiminer}, for instance, uses the frequency of API calls  
to classify malware apps. Recently, MaMaDroid~\cite{MaMaDroid} leverages the similarity in the code of apps as extracted from the transition probabilities between different API calls to build an efficient Android malware detection system.  
However, the increased code complexity and the use of obfuscation techniques as a way to protect apps from code theft, reverse engineering and code inspection, pose several challenges to static analysis of malware apps. 

We set the objective of building an effective mobile malware detection tool that is resistant to code obfuscation techniques. We propose DaDiDroid, a tool that runs static code analysis, leveraging apps' API call graphs to characterize the functional behaviour of apps and filter out malicious behaviour, even in presence of an obfuscated code.  Specifically, DaDiDroid builds, for each Android app, the API families calls graph (e.g., API packages such as  {\it android.util.log} are abstracted to {\it android}) and extracts the weighted directed graphs features (we use 23 unique graph features) to model the Android app behaviour. Using the graph features, we then rely on a classifier to establish whether or not the directed weighted graph belongs to a benign app. %Figure \ref{fig:Scheme} shows an overview of DaDiDroid.    

This paper makes the following contributions:

{\bf Open Source Malware Detection Tool:} We present our open source\footnote{For further research, we plan to release DaDiDroid to the research community.} malware detection tool, DaDiDroid, that leverages features from the weighted directed graph representation of  apps' API calls (\S~\ref{sec:proposedScheme}).

{\bf Efficacy of Graph Features:} We empirically evaluate our proposed system and demonstrate the efficacy of the graph features as peculiar features distinguishing between benign and malicious apps. We use a combination of six publicly available datasets to evaluate the effectiveness of DaDiDroid and determine that, in a clean environment (i.e., no obfuscation used by apps) it detects malware apps with an accuracy of 96.5\%. We present the distributions of some selected graph features to show how they could be used as discriminating features of (obfuscated) malware and benign apps. 

{\bf Comparison against MaMaDroid: } We empirically show (cf. \S~\ref{sec:results} and \S~\ref{sec:obfuscation}) that the accuracy of MaMaDroid, a recent static analysis mobile malware detection tool, is heavily dependent of the balance in the training dataset as captured by the ratio of benign versus malicious apps. In addition, we show that MaMaDroid is prone to obfuscation techniques with a generated false negatives %rate 
as high as 86.3\%.

{\bf DaDiDroid's Robustness Against Obfuscation:} We show that DaDiDroid is robust against several code obfuscation techniques. With no obfuscated apps present in the training set, DaDiDroid accurately classifies 91.2\% of all 
obfuscated apps\footnote{Android apps that use several code obfuscation techniques, further reviewed in \S~\ref{sec:obf_techniques} and discussed in \S~\ref{sec:obfuscation}.}, an average 17.7\% higher success rates when compared to MaMaDroid. 
\vspace{-0.7cm}
\section{\uppercase{Overview of Obfuscation Techniques}}
\label{sec:obf_techniques}
\vspace{-0.45cm}

An Android app package (i.e., APK) comes a compressed file containing all the module and content. Generally, the compressed APK include: four directories ({\it res}, {\it assets}, {\it lib}, and {\it META-INF}) and three  files ({\it AndroidManifest.xml}, {\it classes.dex}, and {\it resources.arsc}). Among these components, {\it classes.dex} contains Java classes of an app, organized in a way the {\it Dalvik} virtual machine can interpret and execute. 

In general, obfuscation attempts to obscure a program 
and makes the source code\footnote{Some obfuscation techniques may also attempt to transform {\it machine code}, however, in this work we consider source code obfuscation techniques.} more difficult for humans to understand, to evade malware detection tools, or both. Developers can deliberately obfuscate code (i.e., Java classes in {\it classes.dex}) to protect app's (malicious) logic or purpose, prevent tampering, or to deter reverse engineering efforts. Common obfuscation techniques used by apps include:

\textbf{Call Indirection:} This involves the manipulation of apps' call graphs by calling any two methods from each others. In essence, in this obfuscation technique, a method invocation is moved into a new method which, in turn, is invoked in place of the original method. 

\textbf{String Encoding and Encryption:} Strings are very common-used data structures in software development. In an obfuscated app, strings could be encrypted to prevent information leakage. Based on cryptographic functions, the original plain texts are replaced by random strings and restore at runtime. As a result, string encryption could effectively hinder hard-coded static scanning.

\textbf{Packing:} In this obfuscation, an APK file is composed of an \textit{origin} APK and a \textit{wrapper} APK. Generally, a wrapper APK launches the origin APK into memory and executes it on Android devices. A wrapper APK may employ encryption to hide (or limit access to) the original APK which are decrypted at runtime thus often evade static code analysis. 

\textbf{Identifier Renaming:}  
In software development, for good readability, code identifiers' names are usually meaningful, though developers may follow different naming rules. However, these meaningful names also accommodate reverse-engineers to understand the code logic and locate the target functions rapidly. Therefore, to reduce the potential information leakage, the identifier's name could be replaced by a meaningless string. Generally, with this obfuscation technique, a given API's name such as {\it myApp.net.myPackage} 
is transformed to at most three-letters words such as {\it a.bc.def}.  
\vspace{-0.45cm}
\section{\uppercase{The DaDiDroid System}}
\label{sec:proposedScheme}
\begin{figure*}[!h]
\centering
\vspace{-0.5cm}
     \includegraphics[width=0.98\textwidth]{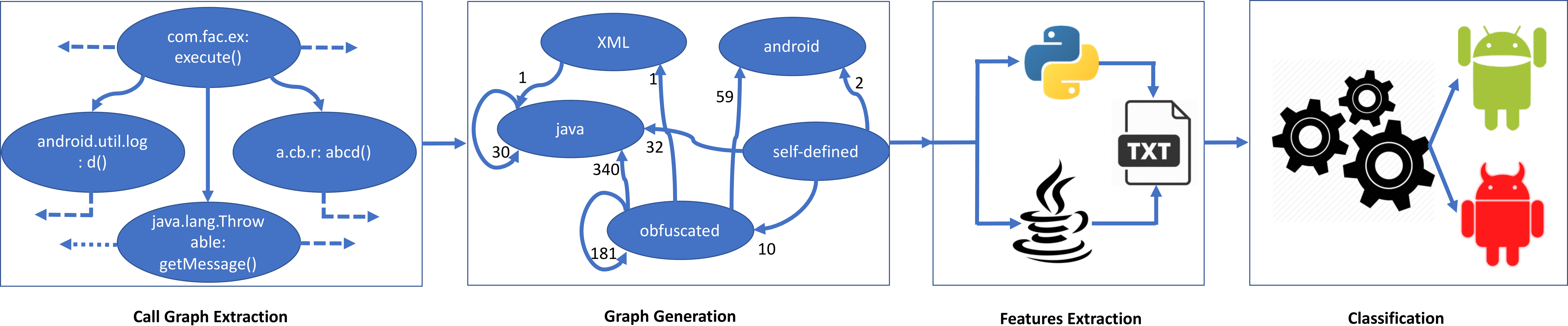}
     \vspace{-0.05cm}
      \caption{Overview of DaDiDroid: The dashed-arrows represent calls to other functions that are not shown in the figure.}% \ik{Does not add much and must be removed.}}
      \vspace{-0.25cm}
      \label{fig:Scheme}
\end{figure*}

\vspace{-0.45cm}
Figure~\ref{fig:Scheme} depicts an overview of the four {\it steps} of the DaDiDroid's malware detection system. By leveraging source code analysis and graph theory, we extract APIs call graphs of Android apps ({\it step} 1) and then model the relationships among APIs ({\it step} 2). We use various graph features ({\it step} 3) and machine learning algorithms ({\it step} 4) to classify apps.  
We provide details of these steps in the following.

{\bf APIs Call Graph Extraction:} We aim to model the behaviour of apps (benign and malware) from the API calls in the source code. To this end, the first step in DaDiDroid is to extract the apps' call\footnote{An Android often requires user's interactions to switch back to app's main GUI. The switching among apps GUIs as programming is termed method callback or simply \textit{callback}. For ease of presentation, we use a generic term, ``call'', to also denote callback.} graphs by performing static analysis on the apps' APKs (Android Apps packages). 
We use Soot~\cite{Soot}, a Java optimization and analysis framework, to extract API call graphs and then instrument FlowDroid (v2.5)~\cite{FlowDroid} to extract dependencies among the APIs. 

Our intuition is that malware may use calls for various operations, in a different graph structure, than benign apps (further elaborated in Section~\ref{subsection:intuition}). To ensure the benign graph structures (resp. graph features such as diameter, in/out degree, see Table~\ref{tabl:metric_summary}) are different enough from the malicious ones, we abstract API calls package to their respective families.

The APIs calls abstractions refers to all possible calls of APIs from every and each API present in the code. That is, for each app in our dataset (see Table~\ref{table:marvin}), we determine all possible calls amongst the APIs. For instance, as depicted in Figure~\ref{fig:Scheme} (step 1), we find that the API call {\it com.fac.ex:Execute()} instantiates the functions {\it android.util.log:d()} and {\it java.lang.Throwable:getMessage()}. 
 For the API family abstraction, we first derive the family names of the APIs and then construct the family call graphs. 
 For example, from {\it android.net.http} and {\it java.sql} APIs we obtain API family names {\it android} and {\it java}, respectively and build the graph of calls based solely based on the family names of the APIs. 
 
The abstraction provides resilience to API changes in the Android framework as families are often added and deprecated less frequently than single API calls. At the same time, this does not abstract away the behavior of an app. 
To implement similar functionalities or operations, families include and assemble classes and interfaces that may be frequently modified by developers. By using API family names for modeling the types of operation, we can limit the consequences of any modification to the included classes or interfaces. For instance, {\tt java.sql} provides in/out and update operations to modify databases. However, {\tt java} family also provides additional classes for other operations. Thus by using abstract family names of APIs, we not only capture the behaviour of a given app but also ensure the resiliency of DaDiDroid against frequent changes to methods or functions of classes and interfaces.
%As families include classes as well as interfaces used to perform similar operations on set of similar objects, we can model the types of operations from the family names independently of modifications, if any, in the underlying classes and interfaces. For instance, we know that the {\it java.sql} package is used for database input, output, and update operations even though there are different classes and interfaces provided by the package for such operations. 
 
Developer may extend the functionalities of their apps by importing a self-defined API. As mentioned earlier, developers also often resort to obfuscate the APIs in the source code. To capture the self-defined and obfuscated APIs in during our call graph extraction process, we rely on previous work~\cite{MaMaDroid} and introduce two types of family names in our APIs calls abstraction: self-defined and obfuscated. 
We rely on previous works~\cite{api_mingling}~\cite{MaMaDroid} to further determine the difference between these two APIs. In principle, we describe an API to be obfuscated if either at least 50\% of its functions or methods are at most \textit{3 characters} (see \S~\ref{sec:obfuscation} for further information) or if we are unable to determine what is implemented, extended, or inherited by its class~\cite{api_mingling, MaMaDroid}. %class implements, extends, or inherits, due to identifier mangling~\cite{api_mingling}.
When operating in API abstraction level, we abstract an API call to its package name using the list of Android packages, which as of Android OS v8.0~\cite{level26index} includes 243 Android and 101 Google APIs. 

DaDiDroid detects if malware developers attempt to define their ``self-defined'' APIs such as {\it com.google.MyMalware} with API names {\it com.google} (or family name {\it com}). The derived lists of Android and Google APIs and classes ensure the detection of any attempt  that malware developers may perform to evade DaDiDroid by naming their self-defined APIs to legal {\it android} and {\it google} APIs such as {\it com.google.MyMalware} which could have been abstracted as {\it com.google} without APIs classes whitelist. 

Overall, for Android API Level 26~\cite{level26index}, we obtain 387 distinct APIs and 12 different families ({\it Android}, {\it dalvik}, {\it java}, {\it javax}, {\it junit}, {\it apache}, {\it json}, {\it com}, {\it xml}, {\it google}, {\it self-defined}, and {\it obfuscated}).
We evaluated DaDiDroid in family and API abstraction modes. We observe that DaDiDroid's results were similar to that of its operation in API abstraction mode. To ease presentation, in \S~\ref{sec:results}, we present our analysis of DaDiDroid operating in family abstraction mode. 

{\bf Graph Generation:} We model the relationships among API calls (resp. API families) in the Android apps as weighted, directed graph $G({\bf V}, {\bf E}, {\bf W})$. As DaDiDroid uses static analysis, the graph obtained from Soot and FlowDroid represents the flow of functions that are potentially called by the app. Each node in ${\bf V}$ corresponds to a unique API call (resp. API family name) in the Android app. A directed link $(u, v) \in {\bf E}$ is presented in the graph if and only if an API, $u\in {\bf V}$, calls a method in another API, $v\in {\bf V}$. A weight $w\in {\bf W}$ represents the number of times $u$ calls a method in $v$. Note that the links $(u, v, w)$ and $(v, u, w)$ may both exist if Android app's APIs are calling each other methods reciprocally. 

For each Android app, we represent the relationships among API calls (resp. API families) as adjacency matrix containing the number of occurrence of possible transitions from $v_i$ to $v_j$. 

{\bf Features Extraction:} We leverage graph metrics (or \textit{features}, listed in Table~\ref{tabl:metric_summary}) to quantify the relationship among API calls. As some of the graph metrics, such as clustering coefficient, are defined only for undirected graphs, we first convert our directed graphs into a weighted, undirected graphs and then extract the corresponding graph metrics. 

{\bf Classification:} The last step in DaDiDroid is to perform classification, i.e., labelling apps as either benign or malware. 
To this end, we use a supervised two-class Support Vector Machine (SVM) classifier~\cite{Muller01anintroduction}, $k$-Nearest Nighbours ($k$-NN)~\cite{knn}, and Random Forest~\cite{Breiman:rf}, all implemented using \emph{Weka}~\cite{hall2009weka}, an open source machine learning library in Java programming language. 

We form two classes by labelling malware and benign apps' graph features (cf. Table~\ref{tabl:metric_summary}) % 
as positives and negatives, respectively. We use 80\% of the instances for training and 20\% for testing. For two-class SVM, appropriate values for parameters $\gamma$ (\emph{radial basis function kernel} parameter~\cite{Scholkopf:2001:ESH:1119748.1119749}) and $\upsilon$ (SVM parameter) are set empirically by performing a greedy grid search on ranges $2^{-10} \leq \gamma \leq 2^{0}$ and $2^{-10} \leq \upsilon \leq 2^{0}$, respectively, on each training dataset. Likewise, for other classifiers, we perform empirical tests to set their parameters. 

\vspace{-0.7cm}
\section{\uppercase{Dataset in Use}}
\label{sec:methodAndDataset}
\vspace{-0.35cm}

{\bf Dataset:} We use previously published datasets which include 63,693 Android apps (\texttt{apk} files) consisting of 43,262 benign and 20,431 malware samples. Table~\ref{table:marvin} summaries the dataset in use. We aim to verify that DaDiDroid is robust to the existence of a variety of Android malware samples as well as to heterogenous APIs. We then consider a dataset that includes a mix of obfuscated and non-obfuscated apps as well as newer and (relatively) older apps with a publication date ranging from August 2010 to June 2018. 
Our dataset consists of samples from the following data sources: 

{\it Marvin~\cite{lindorfer2015marvin}.}  This source consists of a total of 38,740 apps, constituted from 28,181 benign apps and 10,559 malware samples.

 {\it Drebin~\cite{arp2014drebin}.} It consists of 5,525 malware samples collected from August 2010 to October 2012. % 

 {\it Old Benign.}  This dataset includes a sample of 5,879 benign apps collected by PlayDrone~\cite{PlayDrone} between April and November 2013 and published on the Internet Archive Machine~\cite{archivemachine}. % 

{\it New Benign.}  We use a customised crawler and unofficial Google Play API~\cite{unofficialapis} to download a sample of 1,788 free apps, belong to 29 different categories, from the Google Play.  We use VirusTotal to determine whether or not the app's embed in their source codes. %t 

{\it ObData I.}  % 
We obtained this dataset from Garcia et al., ~\cite{lightweight} which consists of obfuscated apps employing various code transformation (i.e., obfuscation) techniques such as Encoding and Call Indirection (explained later in \S~\ref{sec:obfuscation}). % 

{\it ObData II.} Apps developers often minify apps' source code by mapping the API names to names such as \texttt{myApp.net.myPackage} with at most three letters: \texttt{a.b.c} or \texttt{a.bc.def} (cf. \S~\ref{sec:obfuscation}). By parsing the API names that correspond to a given Android app in the Marvin dataset, we construct a sample of 7,975 apps that employ source code minification technique.

%\end{itemize}
\begin{table}[ht!]
  \centering
  %\vspace{-0.5cm}
  %\scriptsize
  \small
 \tabcolsep=0.020cm
  \def\sym#1{\ifmmode^{#1}\else\(^{#1}\)\fi}%
  \begin{tabular}{l*{5}{c}}
    \toprule
\textbf{Dataset} & \textbf{\#Apps} & \textbf{\# Benign}& \textbf{\# Malicious}& \textbf{Date Range}\\
    \midrule
    Marvin~\cite{lindorfer2015marvin}     & 38,740 &  28,181    & 10,559     &  06/12--05/14\\
    Drebin~\cite{arp2014drebin}      & 5,525 &  -    & 5,525     &  08/10--10/12\\
    OldBenign    & 5,909 &  5,909    &  -     &  04/13--11/13\\
     NewBenign    & 1,788 &   1,788   & -    &  06/17--06/17\\  
    ObData I~\cite{lightweight}     & 883 &   -   & 883 & 06/12--05/14\\
    ObData II~\cite{lindorfer2015marvin}    &  7,975    & 5,884   & 2,091    & 05/13--03/14\\
    PackedApps~\cite{dong2018} &  2,873 & 1,500 & 1,373      &  07/16--02/17\\

    \midrule 
    Total &  63,693 &  43,262   &20,431  &  08/10--06/18\\
     \bottomrule
  \end{tabular} 
  
   \caption{Overview of the datasets used in our analysis. We obtain two samples of obfuscated apps: We acquire ``ObData I'' from authors of \cite{lightweight} and use  the Marvin dataset for obfuscated apps (cf. Section~\ref{sec:obfuscation}) to construct ``ObData II''.  }
  \vspace{-0.4cm}
  \label{table:marvin}
\end{table}

{\it PackedApps.} We obtained this dataset from Dong et al.,~\cite{dong2018} which consists of 1,500 and 1,371 \textit{packed}--original APK wrapped in another APK--benign and malicious APKs, respectively.

{\bf Datasets Annotation:} To analyze the classification results of DaDiDroid, we annotate our datasets considering reports from VirusTotal~\cite{virustotal}. VirusTotal is an online service that aggregates the scanning capabilities provided by more than 68 antivirus (AV) tools, %\footnote{Number of AVs as of August 5$^{th}$, 2018},
scanning engines and datasets. It has been widely used in research literature to detect malicious apps, executables, software and domains ~\cite{ikram2016analysis}. After completing the scanning process for a given app, VirusTotal generates a report that indicates which of the participating AV scanning tools detected any malware activity in the app under consideration and the corresponding malware signature (if any). By parsing VirusTotal reports, we extract the number of affiliated AV tools that identified any malware activity and annotate each app in our dataset with the corresponding results.

\vspace{-0.55cm}
\section{\uppercase{Evaluation of DaDiDroid}}% System}}
\label{sec:results}
\vspace{-0.5cm}

In this section, we present a detailed experimental evaluation of DaDiDroid. 

\vspace{-0.15cm}
\subsection{The Intuition}
\label{subsection:intuition}
\vspace{-0.15cm}
We first conduct preliminary experiments in order to verify our main intuition that graph features are valuable to differentiate benign and malicious apps.

\begin{figure}[h!]
\vspace{-0.5cm}
        \begin{center}
                \subfloat[\small Marvin Dataset]{\label{fig:DensityOne}
                        \includegraphics[width=0.47\columnwidth]{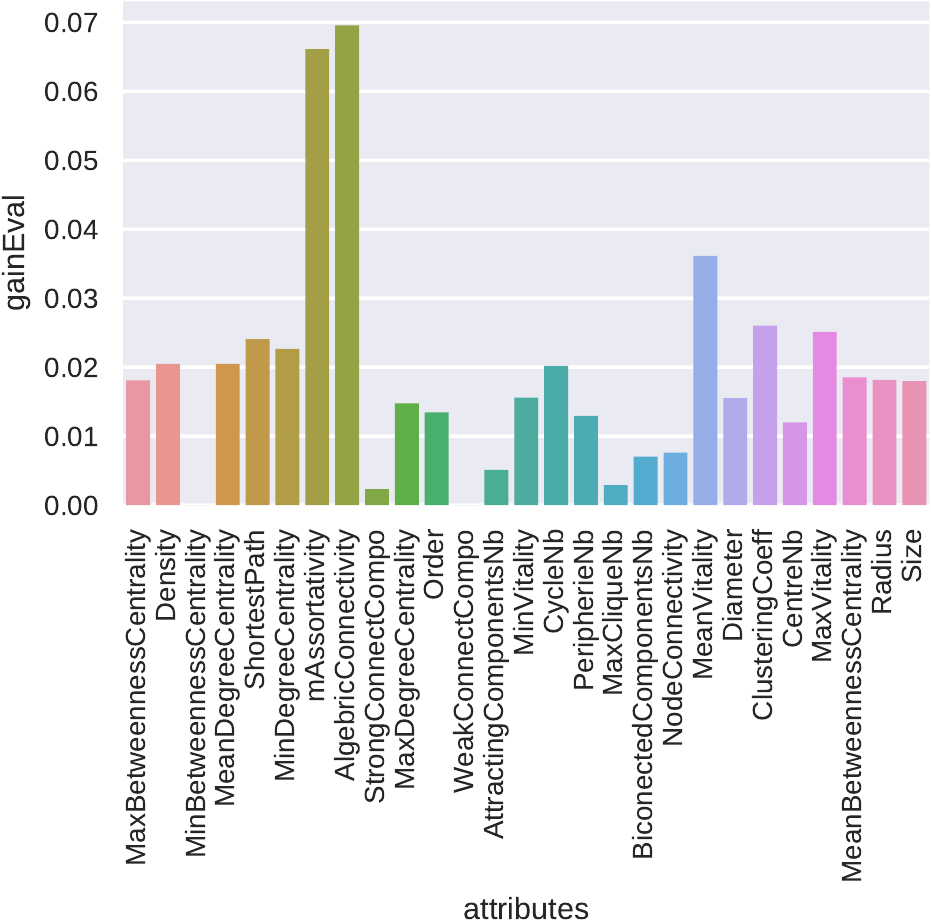}}
%               \quad
                \subfloat[\small Drebin+Oldbenign]{\label{fig:DensityTwo}
                        \includegraphics[width=0.47\columnwidth]{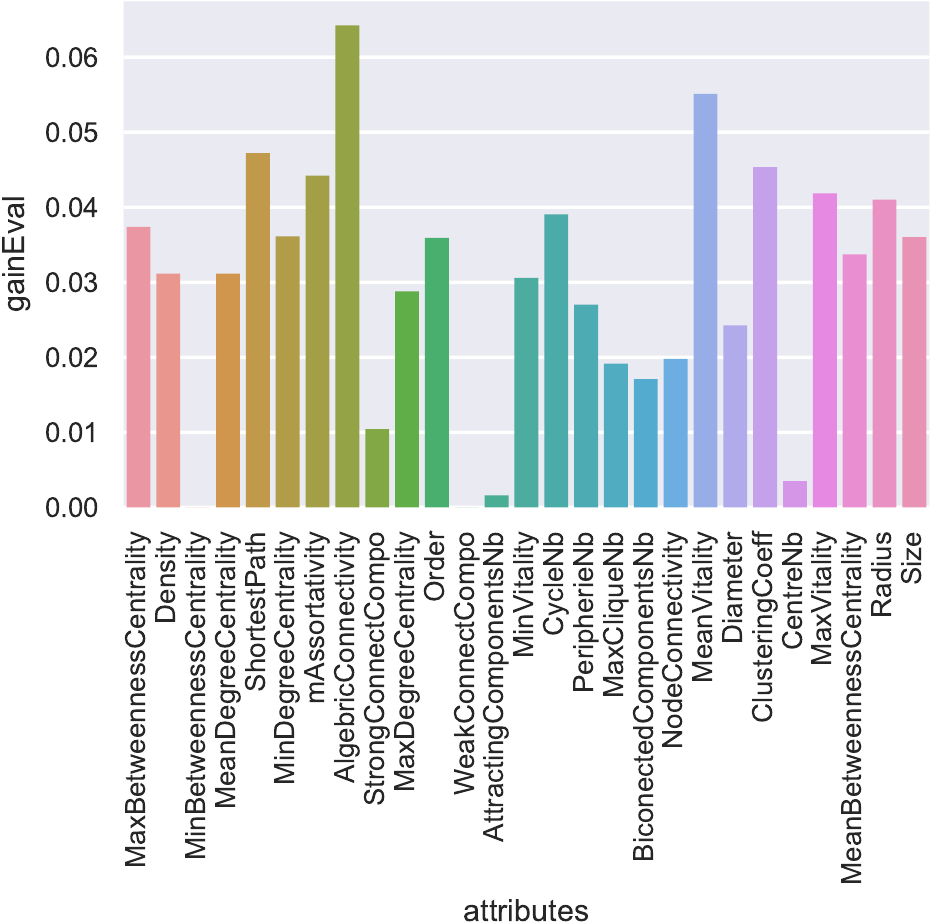}
                }
                \caption{\small Information gain ratio of our graph features for the (a) Marvin dataset and (b) Drebin+Oldbenign dataset. }%\ik{We will discuss this figures in the paper and remove all other in the camera-ready.}}%\ik{TODO: remove '\_' from the x-labels and incline, 30-45 degree, the x-ticks, and normalize y-ticks.}}
                \vspace{-0.45cm}
                \label{feats_contrib}
        \end{center}
\end{figure}

To further analyze the importance of our features in classifying malware and benign apps, we use an information-theoretic metric, \textit{information gain ratio}~\cite{cover2012elements}. This metric is used to quantify the differentiation power of features i.e., our graph metrics or features. In this context, information gain is the mutual information between a given graph metric $X_i$ and the class label $C$ $\in$ \{Benign, Malware\}. Formally, for a given graph feature $X_i$ and the class label $C$, the information gain of $X_i$ with respect to $C$ is defined as:
\begin{equation}
GainRatio(X_i,C)=\frac{(H(C)-H(C|X_i))}{H(X_i)}
\end{equation}
where H($X_i$) = $-\sum_{i \in N}p(x_i)log(p(x_i))$ denotes the marginal entropy of the graph feature $X_i$ and H(C$|$ $X_i$) represents the conditional entropy of class label $C$ given graph feature $X_i$. In other words, information gain ratio quantifies the reduction in the uncertainty of the class label $C$ given that we have the complete knowledge of feature $X_i$.
Figure~\ref{feats_contrib} shows the information gain ratio of our graph features extracted from the Marvin dataset and the Drebin dataset (cf. Table~\ref{table:marvin}). We observe that the ``algebraic connectivity'' and the ``assortativity'' are the most distinct between benign and malicious apps. This analysis reveals that the minimum betweenness centrality (see Figure~\ref{feats_contrib})
and the number of weakly connected components (see Figure~\ref{feats_contrib})
are insignificant in classifying malware and benign apps that correspond to Marvin and Drebin dataset.

%\subsection{Evaluation of DaDiDroid}
\vspace{-0.15cm}
\subsection{Effectiveness of DaDiDroid}
\vspace{-0.15cm}
Using the datasets summarised in Table~\ref{table:marvin}, we perform experiments to analyze the effectiveness of DaDiDroid's classification on benign and malicious samples. % and to investigate the false positive rate and false negative rate of DaDiDroid's classification 
We also study the effect of unbalanced training data on DaDiDroid's performance. %We compare DaDiDroid against MaMaDroid ~\cite{MaMaDroid}. 
We chose to compare to MaMaDroid, as the most recent work in the space, since MaMaDroid also aims to leverage API-calls to extract probabilistic-features building a Hidden Markov model of transitions between API calls. 

% $$F1~Measure=2 \times (\frac{precision~\times~recall}{precision + recall})$$

We use the classic F-measure = $2~ \times$ ($\frac{precision~\times~recall}{precision~+~recall})$, where \textit{precision} = $\frac{TP}{TP+FP}$ and \textit{recall} = $\frac{TP}{TP+FN}$, to evaluate the classification accuracy. TP means \textit{true positives} denoting the number of apps correctly classified as malicious. Likewise, FN (\textit{false negatives}) and FP (\textit{false positives}) indicate the number of samples mistakenly identified as benign and malicious, respectively. For all experiments, we perform 10-fold cross validation using at least one malicious and one benign dataset from Table~\ref{table:marvin}.

\textbf{Effectiveness of Classifiers}: 
In order to generalize the classification of DaDiDroid, we use three machine learning classification algorithms: two-class SVM, $k$-NN, and Random Forest. %

Table~\ref{tab:classifiersAnalysis} summarizes the performance of the classifiers.  Our results show that the Random Forest classifier achieves the highest accuracy with an accuracy of 90\% and 96\% in classifying malware and benign drawn from the Drebin+oldbenign and the Marvin datasets, respectively.  %  
Perhaps due to an overfitting effect in our dataset,  two-class SVM showed poor results and may require further pruning and detailed analysis of parameters, as well as a selection of the optimal set of prominent features to improve the classification performances~\cite{kohavi1995feature, Breiman:rf}. We resort to the Random Forest classifier as our best classifier and use it next to compare against MaMaDroid and in our analysis of the effect of unbalanced training sets as well as the study of the resilience against the obfuscation techniques. 

\begin{table}[h!]
	\tabcolsep=0.05cm
	\scriptsize
	\vspace{-0.2cm}
	\begin{center}
		\begin{tabular}{lrcrrrrrrrrrrr}
			\toprule
	\bf Dataset		& \bf Classifier 	&	\bf  TP	&	\bf  FP	&	\bf Precision &	\bf Recall & \bf FM & \bf Accuracy \\
			\midrule
			
			& RF 	&90.3\% &	 9.7\% &	90.3\%	&	90.3\% &	90.3\% &	90.1 \%\\
			%\hline
Drebin+OB			& SVM 	&70.5\% &	 29.2\% &	71\%	&	70.5\% &	70.4\% &	71.4 \%\\
			%\hline
			& $k$-NN 	&86.8\% &	 13.2\% &	86.8\%	&	86.8\% &	86.8\% &	87.3 \%\\
			\hline
			& RF	&95.7\% &	 6.6\% &	95.7\%	&	95.7\% &	95.7\% &	96.5 \%\\
			%\hline
	Marvin	& SVM 	&81.8\% &	 32.4\% &	81.1\%	&	81.8\% &	81.1\% &	82.7 \%\\
			%\hline
			& $k$-NN 	&94.1\% &	 9.6\% &	94\%	&	94.1\% &	94.1\% &	94.3 \%\\ %\hline
			
\bottomrule
		\end{tabular}
		\caption{\small Performance of our classifiers in classifying malware and benign drawn from Marvin~\cite{lindorfer2015marvin} and Drebin+oldbenign (OB) datasets. Here RM means Random Forest while FM, TP, FP represent F1-Measure, True Positives rate and False Positives rate respectively.} % 
		\label{tab:classifiersAnalysis}
		\vspace{-0.5cm}
	\end{center}
\end{table}

\begin{figure}[!h]
\vspace{-0.5cm}
\begin{center}
\subfloat[\small]{\label{fig:unbalanced_BF}
\includegraphics[width=0.47\columnwidth]{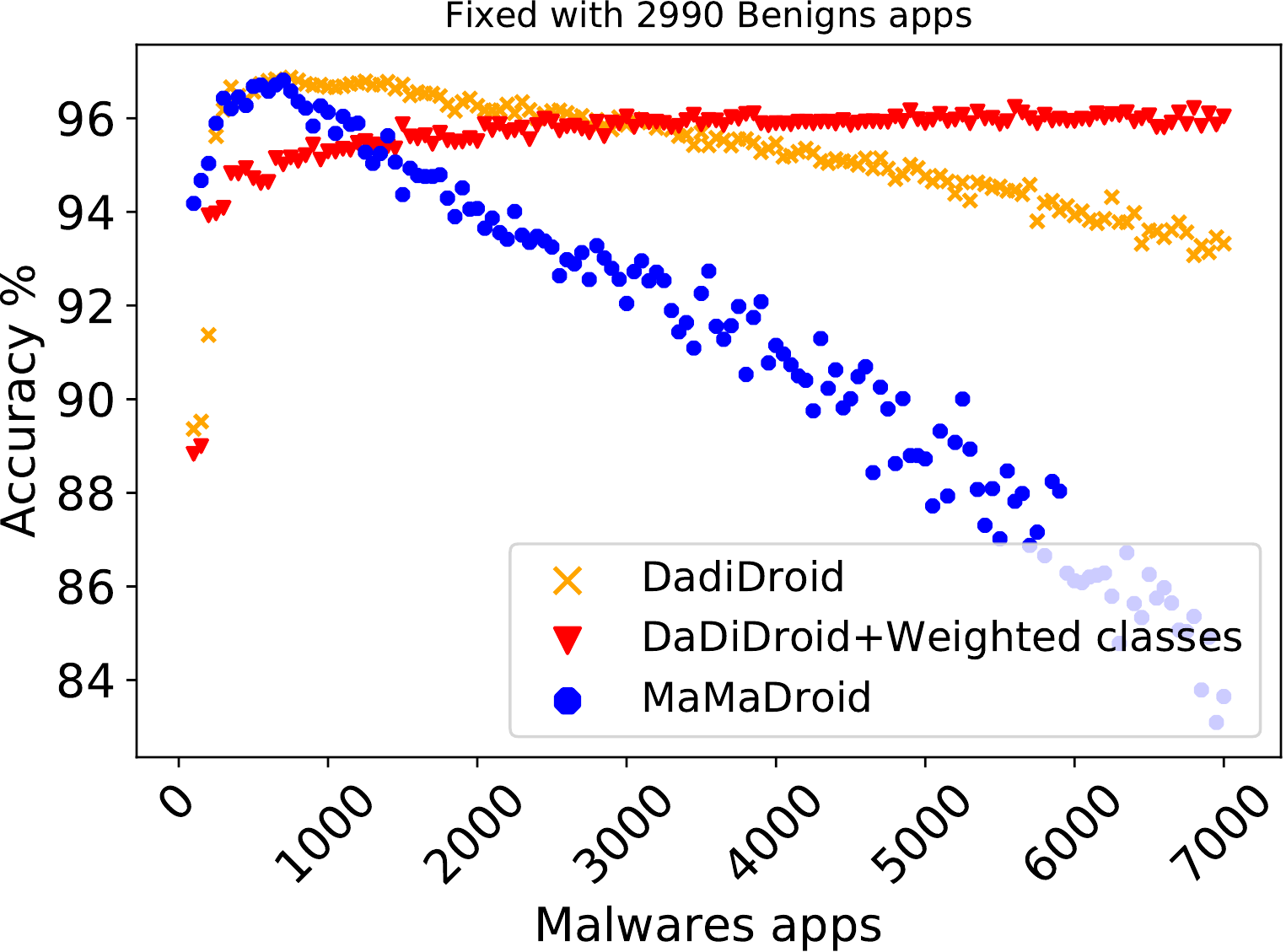}
}%\quad
\subfloat[\small]{\label{fig:unbalancedMF}
\includegraphics[width=0.47\columnwidth]{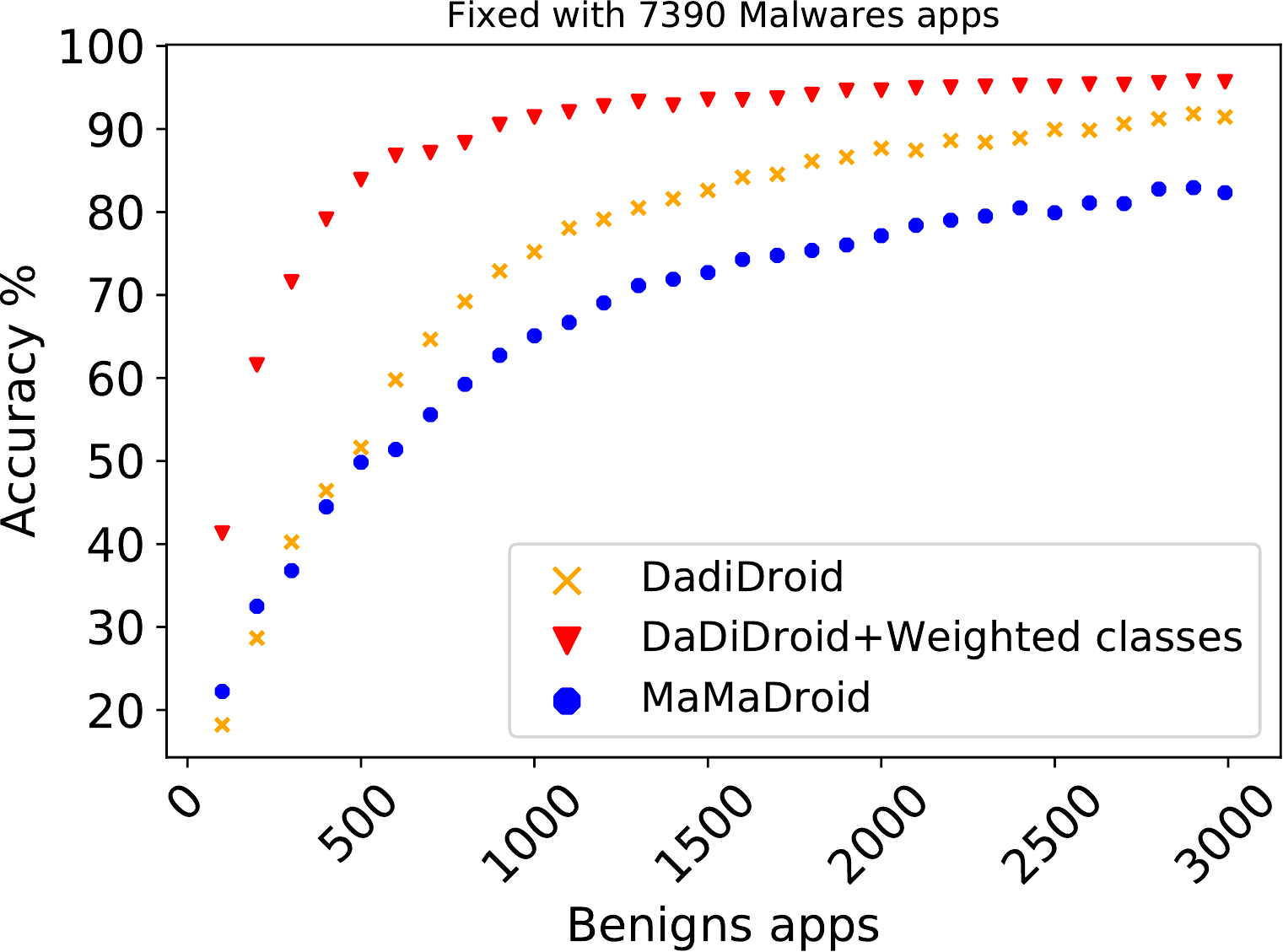}
}
\caption{\small Analyzing the effect of unbalanced dataset on the performance of DaDiDroid (with Random Forest in classification phase) and MaMaDroid: In (a) we use 2,990 benign apps and vary sets of malware apps to measure the \textit{accuracy} while in (b) we use a fixed size of malware apps (7,390) and vary the set of benign apps.}% 
\label{fig:effectofunbalanced_ds}
\end{center}
\end{figure}

\noindent\textbf{Effectiveness of DaDiDroid with Unbalanced Training Dataset:}
In order to determine the minimum number of apps required to train DaDiDroid, we analyze several sets of training and testing datasets drawn from the Marvin dataset (see in Table~\ref{table:marvin}).  
Figure~\ref{fig:effectofunbalanced_ds} depicts the effect of unbalanced dataset on performance of DaDiDroid (with Random Forest). In the case of using fixed size of benign apps and various sets of malware apps (resp. benign apps) in the training phase, we observe that DaDiDroid outperforms MaMaDroid with on average 12\%  higher accuracy. 

Moreover, the results also show that DaDiDroid achieves higher accuracy (96\%) in case of limited number (i.e, 1,500) of benign apps in training phase. In contrast, MaMaDroid \cite{MaMaDroid} shows vulnerability to unbalanced training dataset. In Figure~\ref{fig:unbalancedMF}, in the case of using fixed size of malicious apps and a set 1,500 of benign instances in the training phase, we observe that DaDiDroid outperforms MaMaDroid with on average 26\% higher accuracy.%\ 

\vspace{-0.25cm}
\subsection{Robustness Against Obfuscation Techniques}
\label{sec:obfuscation}
\vspace{-0.25cm}
The main goal of obfuscation is to either secure apps' source codes or to hide apps' malware components. In this section, we evaluate the robustness of our proposed system against several types of obfuscation techniques.  %

\noindent\textbf{Whitespace:} To reduce the readability of source code, this (basic) obfuscation technique adds or removes white-spaces in apps source code. As DaDiDroid leverages API calls (and API family names) only, this obfuscation technique should have no effect on the accuracy of DaDiDroid. %

\noindent\textbf{Call Indirection (CI):} This involves the manipulation of apps' call graphs by calling any two methods from each others. % 
To measure the performance of DaDiDroid to Call Indirection, %(\texttt{t3}), 
we use the Drebin and OldBenign datasets (cf. Table~\ref{table:marvin}) for training our classifiers. We use NewBenign and 210 Call Indirecting malware apps drawn from ObData I~\cite{lightweight} for testing. 
Table~\ref{table:encoding} shows the robustness of DaDiDroid against this type of obfuscation. We observe that 15 (7\%) of apps, having Call Indirection obfuscation, were classified as benign. In contrast, MaMaDroid~\cite{MaMaDroid} mis-classified 63 (30\%) apps that employ Call Indirection obfuscation techniques. 

\noindent\textbf{String Encoding and Encryption (SEE):} With this obfuscation technique, an app's developer encodes the used APIs to unintelligible ones. % }.
To evaluate the performance of DaDiDroid to String Encoding and Encryption,  
we use the Drebin and OldBenign datasets (cf. Table~\ref{table:marvin}) for training our classifiers. We use NewBenign and 673 malware apps, with this obfuscation technique, drawn from ObData I~\cite{lightweight} for testing. 
Table \ref{table:encoding} shows the robustness of DaDiDroid against String Encoding and Encryption obfuscation. We found that 20\% of apps, having String Encoding and Encryption obfuscation, were classified as benign. In contrast, MaMaDroid~\cite{MaMaDroid} results in 86\% of false negatives.

\begin{table}
\centering
\small
\tabcolsep=0.05cm
\resizebox{1\columnwidth}{!}{
\begin{tabular}{r c c c c c c c}
\toprule
{\bf Obf. Tech.}                    & {\bf Scheme}                            & {\bf TP  }                         & {\bf FP    }                     & {\bf P     }                & {\bf R}                       & {\bf FM     }                & {\bf A  }\\

\midrule
                         & \cellcolor[HTML]{EFEFEF}DaDiDroid & \cellcolor[HTML]{EFEFEF}92.9\% & \cellcolor[HTML]{EFEFEF}4.5\% & \cellcolor[HTML]{EFEFEF}70.7\% & \cellcolor[HTML]{EFEFEF}92.9\%   & \cellcolor[HTML]{EFEFEF}80.2\% & \cellcolor[HTML]{EFEFEF}95.2\%\\
\multirow{-2}{*}{CI}     & MaMaDroid~\cite{MaMaDroid}                         & 70.0\%                            & 3.5\%                   & 70.0\%                           & 70.0\%                                                     & 70.0\%                & 93.7\%            \\
\midrule

                         & \cellcolor[HTML]{EFEFEF}DaDiDroid & \cellcolor[HTML]{EFEFEF}79.5\% & \cellcolor[HTML]{EFEFEF}4.5\% & \cellcolor[HTML]{EFEFEF}86.9\% & \cellcolor[HTML]{EFEFEF}79.5\%   & \cellcolor[HTML]{EFEFEF}83.1\% & \cellcolor[HTML]{EFEFEF}91.1\%\\
\multirow{-2}{*}{SEE}     & MaMaDroid~\cite{MaMaDroid}                        & 13.7\%                         & 23.2\%                         & 59.4\%                         & 13.7\%                                               & 22.2\%     & 73.8\%                \\
\midrule
                         & \cellcolor[HTML]{EFEFEF}DaDiDroid & \cellcolor[HTML]{EFEFEF}98.4\% & \cellcolor[HTML]{EFEFEF}0.9\% & \cellcolor[HTML]{EFEFEF}97.9\% & \cellcolor[HTML]{EFEFEF}98.4\%  & \cellcolor[HTML]{EFEFEF}98.2\% & \cellcolor[HTML]{EFEFEF}98.8\%\\
\multirow{-2}{*}{Packing} & MaMaDroid~\cite{MaMaDroid}                         & 98.4\%                         & 1.2\%                        & 97.5\%                         & 98.4\%                                                 & 97.9\%              & 98.7\%            \\
\midrule
                         & \cellcolor[HTML]{EFEFEF}DaDiDroid & \cellcolor[HTML]{EFEFEF}77.1\% & \cellcolor[HTML]{EFEFEF}7.1\% & \cellcolor[HTML]{EFEFEF}91.6\% & \cellcolor[HTML]{EFEFEF}77.1\%   & \cellcolor[HTML]{EFEFEF}83.7\% & \cellcolor[HTML]{EFEFEF}95.2\% \\
\multirow{-2}{*}{IR}     & MaMaDroid~\cite{MaMaDroid}                         & 65.2\%                            & 2.2\%                   & 70.0\%                           & 96.8.0\%                                                    & 77.8\%                   & 65.2\%          \\
\midrule

                         & \cellcolor[HTML]{EFEFEF}DaDiDroid & \cellcolor[HTML]{EFEFEF}82.7\% & \cellcolor[HTML]{EFEFEF}4.5\% & \cellcolor[HTML]{EFEFEF}90.1\% & \cellcolor[HTML]{EFEFEF}82.7\%  & \cellcolor[HTML]{EFEFEF}86.2\% & \cellcolor[HTML]{EFEFEF}91.2\% \\
\multirow{-2}{*}{CI+SEE+Packing+IR}    & MaMaDroid~\cite{MaMaDroid}                         & 27.1\%                        & 3.5\%                         & 79.1\%                         & 27.1\%                                                 & 40.3\%              & 73.5\%             \\

%\hline
\bottomrule                 
\end{tabular}
}
 \caption{Comparison of our proposed system, DaDiDroid, with MaMaDroid on obfuscated datasets (cf. Section~\ref{sec:methodAndDataset}). Here, FM means F-measure while P, R, and A represent precision, recall, and accuracy, respectively.}%, 
 \label{table:encoding}
\end{table}

\noindent\textbf{Packing:} With this obfuscation technique, an app's developer wraps the original APKs inside another APKs. %
To evaluate the performance of DaDiDroid to packing obfuscation, we use 80\% and 20\% of the PackedApps dataset (cf. Table~\ref{table:marvin}) for training and testing our classifiers, respectively, and use 10-fold cross validation in our experiments.
Table \ref{table:encoding} shows the robustness of DaDiDroid against Packing obfuscation. We found that 1.6\% of apps, having Packing obfuscation, were classified as benign. In contrast, MaMaDroid~\cite{MaMaDroid} results in 1.2\% of false positives.

\noindent\textbf{Identifier Renaming:} %This obfuscation technique is similar to encoding with dictionaries but the objective is also to have the smaller package names possible.
With Identifier Renaming, a given API's name such as {\it myApp.net.myPackage} is transformed to at most three-letters words such as {\it a.bc.def}. 
By parsing the API names that correspond to a given Android app in Marvin dataset, we determine this type of obfuscation in our datasets. 
Leveraging previous works~\cite{api_mingling, MaMaDroid}, we describe an API to be obfuscated if either at least 50\% of its functions or methods are at most \textit{3 characters} or if we are unable to determine what is implemented or extended by its class. %cannot tell what its class implements, extends or inherits due to identifier mangling~\cite{api_mingling}.

We draw apps with no Identifier Renaming from Marvin dataset to train MaMaDroid and DaDiDroid and use ObData II (cf.  Table~\ref{tab:classifiersAnalysis}) to measure their robustness against minification. Once again, we use 80\% and 20\% of the data for training and testing, respectively, and use 10-fold cross validation in our experiments. 

\begin{figure}[ht!]
\vspace{-0.5cm}
\subfloat[\small]{\label{fig:minification_F1Acc}
\includegraphics[width=0.49\columnwidth]{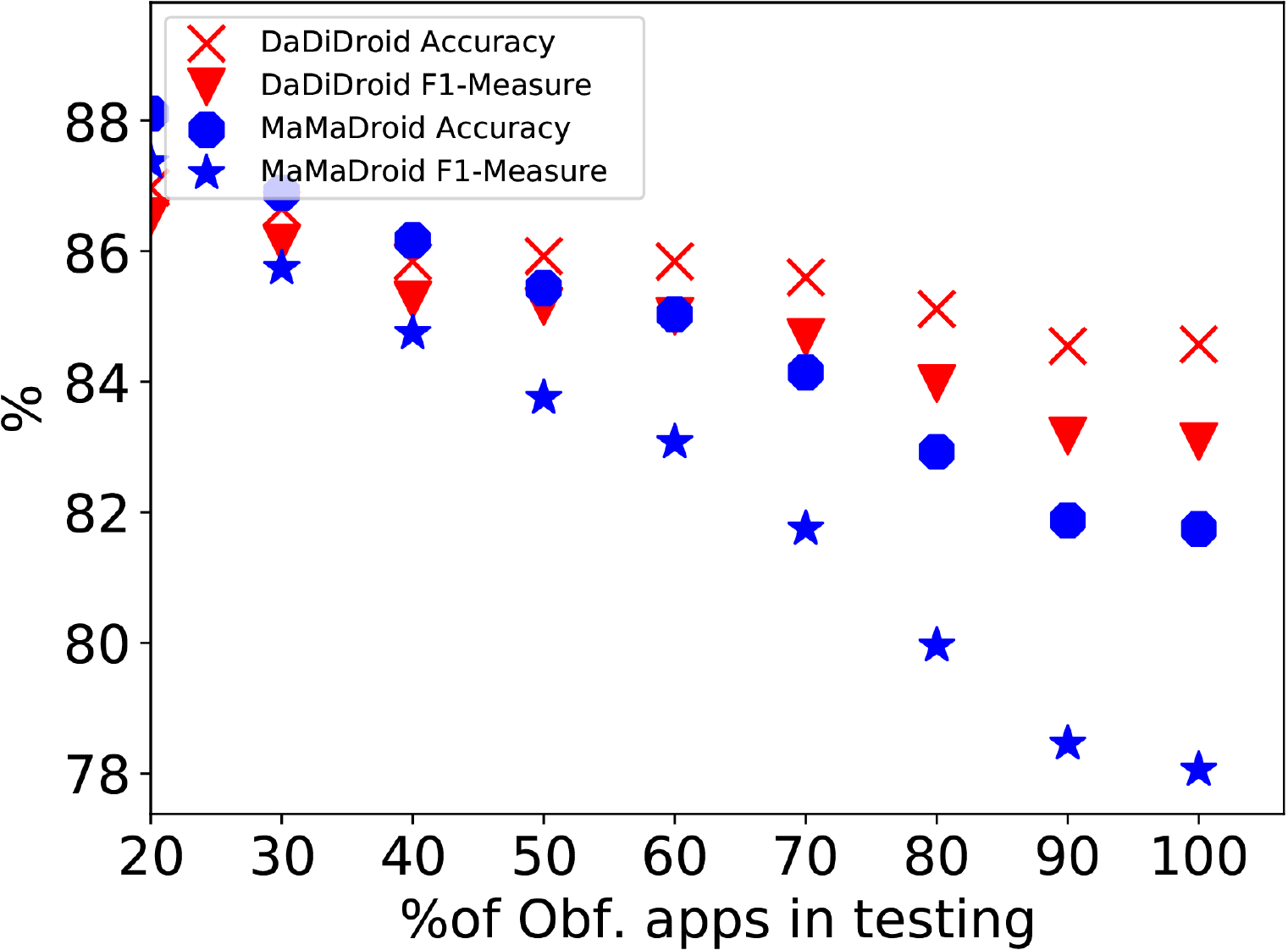}
}%\quad
\subfloat[\small]{\label{fig:minification_FPFN}
\includegraphics[width=0.49\columnwidth]{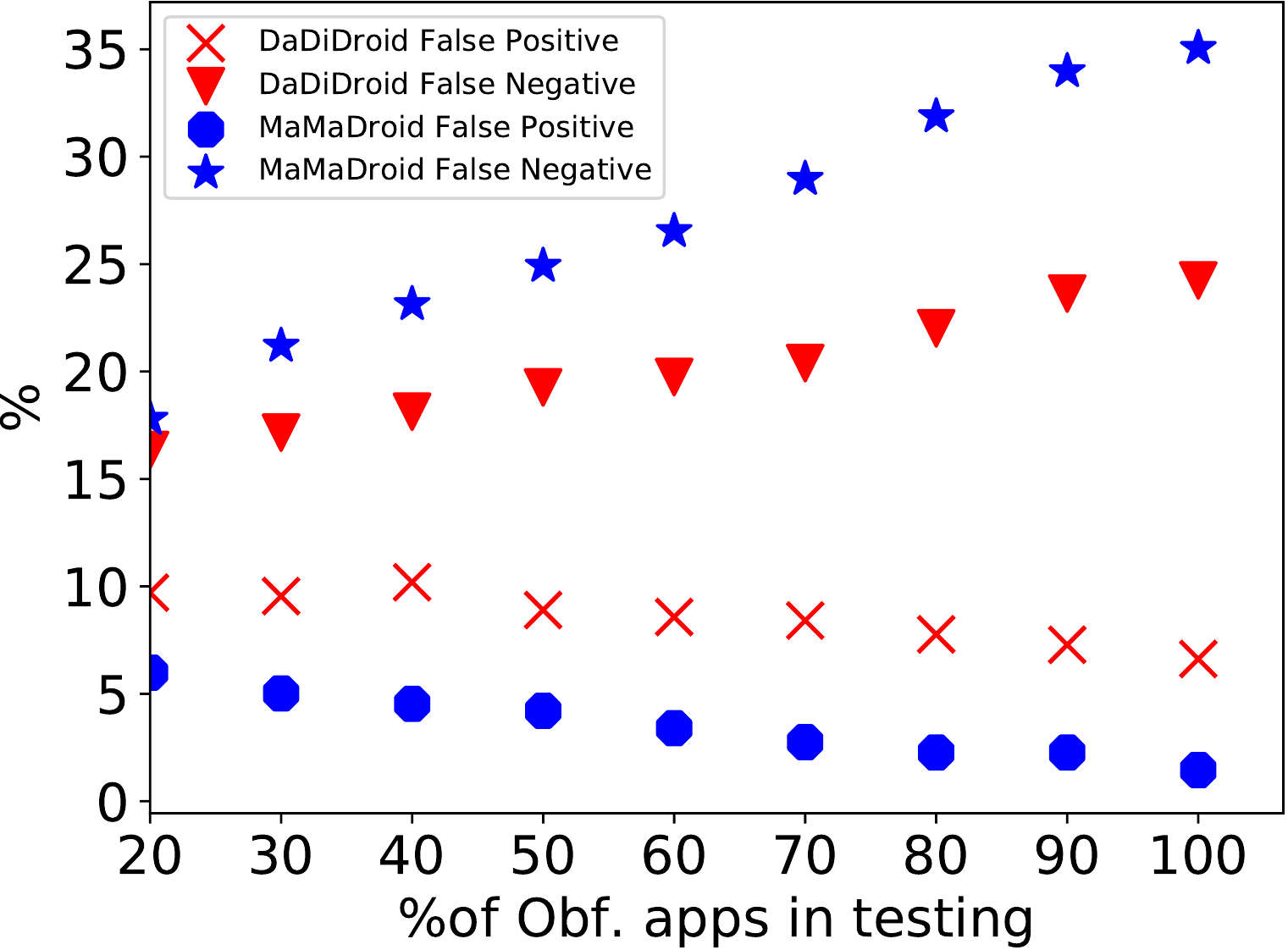}
}
 
\caption{\small Robustness of DaDiDroid against Identifier Renaming (or minification): (a) F1-Measure and accuracy; and (b) false negatives and false positives.} % 
\label{fig:minification_acc}
\vspace{-0.2cm}
\end{figure}

Table~\ref{table:encoding} overviews and Figure~\ref{fig:minification_acc} further illuminates on the robustness of DaDiDroid against Identifier Renaming. With increasing proportions of apps that employ Identifier Renaming, in Figure~\ref{fig:minification_F1Acc}, we observe a decreasing trend in DaDiDroid's and MaMaDroid's performance. When tested with 100\% of ObData II, we observe that DaDiDroid outperforms MaMaDroid: F1-Measure of MaMaDroid drops from 87\% to 78\% whereas we notice merely 5\% (from 88\% to 83\%) decrease in DaDiDroid's F1-Measure. In Figure \ref{fig:minification_FPFN}, we notice that minimification mainly affects the number of false negatives. With 83\% of F1-Measure, DaDiDroid miss-classifies only 20\% of the malware apps in testing dataset as benign. In contrasts, besides its lower F1-Measure (78\%), MaMaDroid results in 35\% of false negatives. % 
Overall, for its high F1-Measure and low false negatives, DaDiDroid pays a very reasonable cost of 7\% of false positives rate.  %

Overall, when tested on samples of benign and malware apps that involve four types obfuscation techniques, we observe (cf. Table~\ref{table:encoding}) that DaDiDroid outperforms MaMaDroid with on average 46.2\% and 17.7\%, respectively, higher F1-Measure and accuracy, however, with a cost of 1\% higher false positive rate than that of MaMaDroid. 

% \vspace{-0.55cm}
\section{\uppercase{Related Work}}
\label{sec:rwork}
\vspace{-0.40cm}
\par A number of studies have characterised and detected malicious activities in Android apps. Several works~\cite{ikram2016analysis, AppFence, taintDroid, continella2017obfuscation}  %burguera2011crowdroid, continella2017obfuscation} 
rely on dynamic runtime-network analysis to detect potential malicious behaviour of Android apps. To reduce the resource required in performing dynamic analysis, a large body of works~\cite{StaticPrivacyLeaks, arp2014drebin,  chen2016stormdroid} %\cite{StaticPrivacyLeaks, arp2014drebin, chen2016stormdroid, aafer2013droidapiminer, chen2016stormdroid} 
perform static code analysis to detect malware component in Android apps. However, due recent development in obfuscation techniques and malware developers' affinity to obfuscate their apps~\cite{dong2018}, static analysis of Android malware apps is an active research topic.
CrowDroid \cite{burguera2011crowdroid} performed dynamic analysis apps network traces to detect malware activities of Android apps. By collecting sequence of requests sent to remote servers, it built features vectors that consisted of number of calls of a specific API to differentiate malware apps from benign apps. 
Drebin~\cite{arp2014drebin} performed static analysis of API calls and requests to permission. It also analyzed hard-coded URLs in apps' source codes to determine apps' benign or malicious activities. DroidApiMiner~\cite{aafer2013droidapiminer} proposed a new approach by using the frequencies of API calls to sensitive APIs. StormDroid~\cite{chen2016stormdroid} improved the detection of DroidApiMiner by analyzing APIs calls sequences along with the API calls frequencies analysis. 
\par Similarly, MaMaDroid~\cite{MaMaDroid} improved the in-efficiency of both DroidAPIMiner and StormDroid~\cite{chen2016stormdroid} by using the transitions probabilities among API calls as features. However, we empirically analyzed that MaMaDroid is dependent on the type of data used for training (i.e., apps categories, ratio benign on malicious apps, and apps release dates) and is prone to obfuscation techniques (see Section~\ref{sec:obfuscation}) employed by malware apps. Overall, these techniques focused on sensitive API calls and privileged permissions that often lead to high false positives: a large number of benign apps often use sensitive API calls and request permissions for certain functionalities thus they are wrongly classified as malware.  Moreover, these solutions seldom detect obfuscated malware components in apps, resulting in poor performance. 
Zhang et al.,~\cite{ClassifWeightedDirectGraph} used weighted graph models, extracted from Android apps' codes semantics, to quantify the similarities among malware apps. In contrast, we used metrics derived from weight directed graphs to differentiate among benign and malware apps. To analyze malicious software in desktop computing platform, GZero~\cite{GraphApproachMalwares} collected the call sequences of software executables to build graph theoretic features. In the same spirit, in this work, we extracted an extended set of graph features and used showed the efficacy of graph theoretic metrics in classifying benign and malware apps. Moreover, we evaluate the robustness of our proposed scheme against unbalanced datasets as well as shown its resiliency against several type obfuscation techniques.

\vspace{-0.55cm}
\section{\uppercase{Conclusion}}
\label{sec:conclusion}
\vspace{-0.35cm}
We presented DaDiDroid, an obfuscation resilient tool to detect Android malware based on modeling the API calls as a weighted, directed graph. We evaluated the effectiveness of DaDiDroid with several datasets and tested its robustness against different types of obfuscations. We shown that DaDiDroid classify up to 96.5\% of benign and malicious apps. In worst case, when no obfuscated apps in training set and testing set contain all obfuscated apps (i.e., apps ``Encoding'' obfuscation technique), DaDiDroid achieved 91.2\% of accuracy with only 17.3\% of false negative rate and 4.5 false positive rate. Moreover, DaDiDroid use small features vectors which can be customized with more or less features from the graph theory and even the extraction process depending on the context of the classification.
We compared DaDiDroid to MaMaDroid~\cite{MaMaDroid}, a Markov chain modeling tool for Android apps, showing that beside achieving similar high results on common datasets, DaDiDroid is much more resilient to changes in the datasets and especially to obfuscation techniques.  % 
In the future we plan to improve DaDiDroid by complementing features derived from dynamic runtime analysis of apps. %

\subsubsection*{Acknowledgments}

This work was partially supported by research grants from Macquarie University Deputy Vice-Chancellor Research (MQ-DVCR) and Optus Macquarie University Cyber Security Hub (OMUCSH).  Any opinions, findings, and conclusions or recommendations expressed in this material are those of the authors or originators and do not necessarily reflect the views of the MQ-DVCR or of the OMUCSH. The authors would like to thank Minhui (Jason) Xue and the anonymous reviewers for their constructive feedback on the preparation of the final version of this paper. 

\scriptsize
%\small
%\bibliographystyle{apalike}
\vspace{-0.55cm}
\bibliographystyle{abbrv}
\bibliography{DaDiDroid}
\vspace{-0.45cm}

\begin{table*}[h!]

\centering
%\begin{longtable}{lp{12.5cm}}
\small
%\scriptsize
\vspace{-2.45cm}
\resizebox{1\textwidth}{!}{
\begin{tabular}{lp{12cm}}
%\begin{center}
\toprule
%Category              & Metric & Definition \\
Feature & Description \\
\midrule

Graph Size            &  Denotes number of edges (i.e., relationships) of a node (i.e., API or family) in a graph.\\
Graph Order & Corresponds to the number nodes in a graph. \\
Number of Cycles & %The number of simple cycles (elementary circuits) of the directed graph. 
A cycle is a path of edges and vertices wherein a vertex is reachable from itself.  \\
Degree Assortativity Coefficient &    In a Directed graph, in-assortativity and out-assortativity measure the tendencies of nodes to connect with other nodes that have similar in and out degrees as themselves, respectively. To compute this feature we use the Pearson correlation approach which is way faster and provide representative results. Positive values of assortivity coefficient, $r$, indicates a correlation between nodes of similar degree, while negative values indicate relationships between nodes of different degree. In general, $r$ lies between $-1$ and $1$. When $r$ = $1$, the network is said to have \textit{perfect assortative mixing patterns}, when $r$ = $0$ the network is \textit{non-assortative}, while at $r$ = $-1$ the network is completely \textit{disassortative}.\\
Strongly Connected Components &   The number of strongly connected components. A vertex $v_i$ is said to be strongly connected if it is reachable from every other vertex $v_k$. \\
Weakly Connected Components &    A weakly connected component is a maximal subgraph of a directed graph such that for every pair of vertices $(u,v)$ in the subgraph, there is an undirected path from $u$ to $v$ and a directed path from $v$ to $u$. \\
Node Connectivity & The node connectivity is the minimum number of nodes that must be removed to disconnect $G$. \\
Avg. Shortest Path Length &  Average shortest path length is a concept in network topology that is defined as the average number of steps along the shortest paths for all possible pairs of network nodes. \\
Degree Centrality &   The degree centrality a vertex $v_i$ is the fraction of nodes it is connected to.  \\
Betweenness Centrality & Measures the fraction of all pair shortest paths, except those originating or terminating at it, that pass through it, $\frac{2I(P_{jk}, i)}{|V|(|V|-1)}$, where $P_{jk}$ denote the shortest path from vertex $v_j$ to vertex $v_k$,  $P_{jk}$ = $(v_j,v_l,v_m,v_n,...,v_k)$ and $I(P_{jk},i) = 1$ if $v_i$ $\in$ $P_{jk}$ otherwise $I(P_{jk},i) = 0$ when $v_i$ $P_{jk}$. \\
Vitality & Closeness vitality of a node is the change in the sum of distances between all node pairs when excluding that node. \\
Attracting Components & An attracting component in a directed graph G is a strongly connected component with the property that a random walker on the graph will never leave the component, once it enters the component.\\
Graph Density &   It measures number of edges of graph is close to the maximum number of edges.\\
Clustering coefficient & It is a measure of the degree to which nodes in a graph tend to cluster together.\\
Algebraic connectivity & The algebraic connectivity (also known as Fiedler value or Fiedler eigenvalue) of a graph $G$ is the second-smallest eigenvalue of the Laplacian matrix of $G$. This eigenvalue is greater than 0 if and only if $G$ is a connected graph. This is a corollary to the fact that the number of times 0 appears as an eigenvalue in the Laplacian is the number of connected components in the graph.\\
Clique Number &    It corresponds to the size (number of vertices $V$) of the largest clique of the graph. A clique $C$ in an undirected graph is a subset of the vertices such that every two distinct vertices are adjacent.\\
Biconnected Components &  They are maximal subgraphs such that the removal of a node (and all edges incident on that node) will not disconnect the subgraph. \\
Diameter &   The diameter of a graph is the maximum eccentricity of any vertex in the graph. The eccentricity $\epsilon(v)$ of a graph vertex $v_i$ in a connected graph $G$ is the maximum graph distance between $v_i$ and any other vertex $v_j$ of $G$. \\
Radius & The radius $r$ of a graph $G$ is the minimum eccentricity of any vertex.\\
Center Number &  The center is the set of nodes with eccentricity equal to radius, $r$.\\
Periphery Number & The periphery is the set of nodes with eccentricity equal to the diameter.\\

\bottomrule
\end{tabular}
}
\caption{List of our graph metrics (or {\it features}) to quantify the relationship among API calls as well API families.}% (see \S~\ref{subsection:featuresExtraction} for further details).}
\label{tabl:metric_summary}
\end{table*} 
\end{document}